%% file: main.tex
\documentclass[runningheads]{llncs}

\usepackage[utf8]{inputenc}
\usepackage[margin=1.3in]{geometry}
\usepackage{amsthm}
\usepackage{amsmath}
\usepackage{tikz}
\usepackage{relsize}
\usepackage{listings}
\usepackage{algorithm}
\usepackage{algpseudocode}
\usepackage{tikz}
\usepackage{float}
\usepackage{tabularx}
\usepackage{enumitem}
\usepackage{fancyvrb}
\usepackage{xcolor}
\usepackage[sorting=none]{biblatex}
\usepackage{hyperref}
\hypersetup{colorlinks,allcolors=black}

\bibliography{bibfile}

\usetikzlibrary{automata}
\makeatletter
\newcommand{\brokenline}[2][t]{\parbox[#1]{\dimexpr\linewidth-\ALG@thistlm}{\strut\raggedright #2\strut}}
\makeatother

\begin{document}
\title{Formal Verification of Safety Properties Using Interpolation and $k$-induction}
\author{Tephilla Prince\inst{1}\and
Atif Abdur Rahman\inst{2}\and
S. Sheerazuddin\inst{3}} 
\institute{Indian Institute of Technology Dharwad, India \\
\email{tephilla.prince.18@iitdh.ac.in}\and
NSUT East Campus (Formerly AIACTR), India\\
\email{atifrahmanmi1223@gmail.com}\and
National Institute of Technology Calicut, India \\
\email{sheeraz@nitc.ac.in}
} 
\maketitle   

%\begin{center}
%    \huge{Formal Verification of Safety Properties Using Interpolation and $k$-induction}\\[40pt]
%\end{center}
%\rule{\textwidth}{0.5pt}

\begin{abstract}
    \noindent %The interpolation technique can be used to stretch the bounded model checking to the unbounded case and can be used for verification of safety properties.
    This technical report presents implementation of two symbolic model checking algorithms that use SAT/SMT Solvers, namely interpolation based model checking and $k$-induction based model checking. We also do a comparative analysis of these two model checking algorithms.
\end{abstract}
%\rule{\textwidth}{0.5pt}
\section{Introduction}
In any software development process, testing and validation remains an important step before releasing the final product. Formal verification methods are designed to find bugs and check performance with specifications in the system. With a given finite state machine description and a temporal logic property, we can automatically check whether a given property holds or not using model checking. 
Model checking is a method for formally verifying the correctness of finite state systems. It provides counter examples if the system does not satisfy the given specification. %Model checking tools face a common problem known as state space explosion. If $\phi=f(x_1,x_2,\cdots x_n)$, is a formula containing $n$ variables, then there are $2^n$ possible assignments to the variables $x_1,x_2,\cdots x_n$, this is exponential. This is the state explosion problem. For example, if NuSMV program has $10$ Boolean variables, then there are $2 ^ {10}$ possible assignments. There are various approaches to reduce this problem.

Bounded model checking (BMC) is a formal method based approach that may be used to find bugs in a system given as a finite state machine. In BMC, we unfold the transition relation of the finite state machine for $k$ steps and check if there is any property violation. After every iteration, it increases the bound to search for bugs in longer traces. For each $k$, a Boolean formula is constructed which return sat, if there is a counterexample of length $k$. We can check the satisfiability of Boolean formulas using SAT solvers. 

The main problem in BMC is that the property validation can only be checked up to $k$ steps, to check for longer traces we have to increment the $k$. A complete bound can be calculated based on the longest loop-free path between two states. But the longest loop-free path can be exponentially larger than the reachable diameter of the state space. The calculation of such an upper bound can be computationally expensive. %So, the crux is that we have to extend the BMC so that it could verify safety and liveness properties towards the infinite state systems.     

Interpolation is an efficient methodology to enhance BMC to the unbounded case. The concept of interpolation in model checking was first introduced by K.C. Mcmillan in 2003\cite{mcmillan2003interpolation}. In contrast to BMC, here we do not have to compute an upper bound. In BMC, the longest loop free path can become exponentially larger than the reachable diameter but in interpolation-based model checking FSM can unfold upto the diameter of state space. Generally, it terminates before it reaches the diameter. The algorithm and its implementation are discussed in detail in the next sections.

\section{Logic}
Logic allows us to reason about mathematical models for which it has to follow a set of guidelines. Based on the differences in guidelines (syntax and semantics), there are different types of logic. One such type is Modal Logic. It is a specific type of Zero Order Logic. It is the logic of possibility and necessity, past and future, knowledge and belief, and dynamic change. It is developed to study different modes of truth. For different types of modals there exists different types of modal logic\cite{chagrov1997modal}. One of the most studied modal logic is tense logic (temporal logic) or logic of time. Temporal logic allows us to make deductive statements about what was, what has always been and what always will be.

Propositional logic and first-order logic do not have a time component, therefore in some cases, we cannot apply this logic directly. For example, if you want to specify the system functionality it requires a notion of time. Temporal logic is a special kind of modal logic where truth values of assertion vary with time\cite{rescher2012temporal}. It is used in the verification of sequential circuits. An automation system is also a sequential circuit because it depends on the previous and next states. So, we need temporal logic for specifying behavior. Potential models of temporal logic are sequences if states (paths) and property can hold in some states and does not hold in other states. One such model is the Kripke structure which is very similar to a finite state machine.

In Linear-time Temporal Logic (LTL), we use two fundamental temporal operators to construct more complex formulas from simpler formulas. These operators are as follows:
\begin{itemize}[topsep=1pt,itemsep=1pt,partopsep=1pt, parsep=1pt]
\item Next operator  :-  $X\alpha$ - property holds in the next state.
\item Until operator :- \text{$\alpha U \beta$}  - $\alpha$ holds in all states until $\beta$ holds.
\end{itemize}
Also, the LTL language has two derived temporal operators which are quite useful.
\begin{itemize}[topsep=1pt,itemsep=1pt,partopsep=1pt, parsep=1pt]
\item Future operator :- $F\alpha$ - $\alpha$ holds in some future states.
\item Always operator :- $G\alpha$ - $\alpha$ holds in all states.
\end{itemize}
The set of LTL formulas over atomic propositions (AP) is defined as follows:
\begin{itemize}[topsep=4pt,itemsep=2pt,partopsep=2pt, parsep=2pt]
\item If $p \in AP$,  then $p$ is an $LTL$ formula.
\item If $\alpha$ and $\beta$ are $LTL$ formulas then $\neg \alpha\mid  \alpha \lor \beta \mid \alpha \land \beta \mid X \alpha \mid  \alpha  U  \beta$ are $LTL$ formulas.  
\end{itemize}
Temporal logics include interval temporal logic, linear-time temporal logic, computation tree logic, and many other types of logic. The most convenient and common formalism used for specifying properties is an extension of $G$, $F$, and $X$ expressions which are called linear-time temporal logic (LTL). This system only contains operators that deal with the future namely $G$, $X$, and $U$. It is a subset of more complex CTL*, which allows branching. One application of LTL is that it can express safety and liveness properties For more information about automata-based LTL model checking, visit~\cite{Vardi87}. Safety and liveness properties are discussed in the next sections.

\subsection{Safety Properties}
In this section, we formally define the notion of a property and, also, the notions of safety and liveness properties. In this report, we focus on safety properties and their model checking. It has been argued elsewhere that model checking safety properties is computationally cheaper than model checking liveness properties.
\begin{definition}
Given a set of atomic propositions ($AP$), a property over $AP$ is a subset of infinite words over $2^{AP}$.
\end{definition}
Here is an example of a property:
\begin{example}
$p_1$ is true at least once and $p_2$ is always true.
\[\{A_0 A_ 1 A_ 2 \cdots  \mid \forall i, A_i \subseteq AP ~\mbox{and}~ \exists i, p_1 \in A_i ~\mbox{and}~ \forall j, p_2 \in A_j\}.\]
\end{example}
Correctness of a property is always expressed in terms of safety and liveness properties. Any property can be expressed as a conjunction of safety and liveness property \cite{Alpern87}. A characteristic of safety property is that something bad is absent which means it is used to verify that something bad never happens. $P$ is a safety property if there exists a set of bad prefixes such that $P$ is the set of all words that do not start with a bad prefix.

A Property $P$ is a safety property if:-
\begin{itemize}[topsep=0pt,itemsep=1pt,partopsep=1pt, parsep=1pt]
\item  Given any execution Ex, $P$(Ex) is false.
\item  There exists a prefix of $F$, such that every extension that prefix gives an   execution   $F$ such that $P$($F$) is false. 
\end{itemize}

\begin{example}
Always: if $p_1$ is true, then in the next step $p_2$ is true.\linebreak
\hspace*{2cm}this property can be written as\textbf{ G($p_1 \implies Xp_2$)}
\end{example}

\subsection{Liveness}
Liveness property asserts that something good eventually happens whereas safety property asserts that something bad is absent. Liveness property can not be violated in a finite time because something good can happen later whereas safety property can only be violated in a finite time. But liveness property can be satisfied in a finite time contrary to that safety property can be satisfied in infinite time. Liveness is concerned with a program eventually reaching a good state.  We can describe the liveness property by combining the $G$ and $F$ operator:-
\begin{center}
    \textbf{$GFp$} : infinitely often $p$
\end{center}

\section{Bounded Model Checking}
The main problem in model checking is the capacity issue that arises mainly due to the size of the state space. BDDs \cite{andersen1997introduction} can represent the state transition relation in canonical form and we can efficiently traverse the state space to check whether a given property holds or not. Typically, for a large number of variables, it is not feasible for us to construct a BDD because that may require a sufficiently large amount of memory. For linear increases in variables, BDDs grow exponentially \cite{edelkamp2008limits}. In this section, we look at another approach to tackle this problem, namely Bounded model checking (BMC).

The key idea of bounded model checking is that it looks for a counter-example path of increasing length $k$. We have to increase the bound to search for longer traces\cite{biere2009bounded}.
For each $k$, it builds a Boolean formula that is satisfiable if there is a counterexample of length $k$. So, as you can see BMC is more oriented toward finding bugs. The satisfiability of Boolean formulas is checked using SAT procedure and according to that, it returns a satisfying assignment. As we have to increase the bound successively, the correctness of the property can be verified in that limited bound. We can calculate a complete bound to check safety properties based on the longest loop-free path between the states. Unfortunately, the longest loop-free path can be exponentially larger than the reachable diameter of the state space.

For BMC we need the symbolic representation of a system $M=(I,T)$, where Boolean formulas $I$ and $T$ describe the initial states and transition relation respectively. For a given safety property $P$, we can check whether $P$ holds for all states at a distance of $k$ by constructing a Boolean formula $F_k$ with following three clauses:
\begin{description}
\item[1.] $I(S_0)$
\item[2.] $\bigwedge\limits_{i = 0}^{k - 1} P(S_i, S_{i+1})$
\item[3.] $\lnot P(S_k)$
\end{description}
If $F_k$ is satisfiable then it means that some state in $S_k$ (at a distance of $k$ from some initial state in $S_0$) is unsafe (does not satisfy safety property $P$) and we may terminate the procedure. If $F_k$ is unsatisfiable we can increment $k$ and search for longer traces. If $P$ holds for all reachable states then we say that $M$ is $P$-safe.

BMC in its original form can be used for falsification. For completeness, it requires an upper bound. Unfortunately, the calculation of such an upper bound can be computationally expensive. It is important to know if the system is $P$-safe and when to terminate the program or stop incrementing $k$.This property is called completeness. There are several techniques for achieving completeness. Here we discuss the properties such as $Fp$ and $Gp$. In the next section, we introduce how completeness can be achieved with $k$-induction.

The bound where we can stop our BMC algorithm and conclude that the system is $P$-safe is called the completeness threshold (CT)\cite{clarke2004completeness}. So for $Gp$ formulas, the completeness threshold is the minimum number of steps required to reach all states which is also called reachability diameter. We can also use the largest distance of any reachable state from an initial state like CT. Unfortunately, computation of diameter can be expensive. Instead of computing the diameter, we can compute the over-approximation of reachable diameter which is called recurrence diameter, which is just the longest loop-free path from an initial state. As every shortest path is a loop-free path, we can say recurrence diameter is an over-approximation of the reachability diameter.

\subsection{$k$-induction}
We can also use $k$-induction for completeness of BMC, with induction we can check safety properties. In this algorithm, first, the base case is checked if it is unsatisfiable then only we move forward towards the induction step. The base case for $k$-induction is:
$I(S_0) \land \neg P(S_0)$.
%    \text{I($S_0$) \hspace{5pt} $\land$ \hspace{5pt} $\land_{i = 0\cdots k - 1 }$ T($S_i$, $S_{i+1}$) \hspace{5pt} $\land$ \hspace{5pt} $\land$ $\neg$ $P_k$}

If the result is sat then we can simply abort the program and conclude that input FSM is unsafe at the initial state. Otherwise, if the result is unsat, we can proceed further toward the induction step. Also, we can conclude that all initial states are P-safe for $k$ = 0.

The induction formula which is to be checked with every increment of $k$ is:
\[\bigwedge\limits_{i=0}^{k-1} T(S_i, S_{i+1})\land \neg P(S_k)\]
For base case, $k$ = 0 if there is no counter example, it increments the $k$ and checks for the next step. For $k = 1$ it checks for \text{T($S_0$, $S_1$) $\land \neg$ $P(S_1)$} and if it returns satisfiable then input FSM is unsafe for $k = 1$ and we abort the program. Otherwise, we keep on incrementing $k$ and check for every iteration. For example, for $k = 2$, it checks for the formula \text{T($S_0$, $S_1$) $\land$ T($S_1$, $S_2$) $\land \neg$ $P(S_2)$} and do the same checks. 

How to know when it is safe to stop incrementing and conclude that the system is $P$-safe? We can not wait for \text{I($S_0$) $\land$ T($S_0$,$S_1$) $\cdots$ $\land$ T($S_{i-1}$, $S_i$)} to become a contradiction because given that there is an initial state, this will never happen because we assume that every state has successor through $T$, so there will always be loops in state space. Instead, we can check the following:\begin{center}
    
 \text{I($S_0$) $\land$ loop-free($S_0$,$S_1$) $\cdots$ $\land$ loopfree($S_{i - 1}$,$S_i$)}\end{center}
 and if there is a contradiction then we can safely conclude that the system is $P$-safe. 

\begin{algorithm}%[H]
\caption{BMC with $k$-induction}\label{alg:kind}
\begin{algorithmic}
\Statex $V$ = [ ] \algorithmiccomment{List $V$ holds the variables used in the runs} 
\Statex $Run$ = [ ] \algorithmiccomment{List $run$ holds the expressions $run_0$, $run_1$, etc.}
\Statex $Path$ = [ ] \algorithmiccomment{List $path$ holds the expressions $path_0$, $path_1$, etc}
\Statex $Run_0$ = $I$, $subp_0$ = $P$

\If{(\text{$Run_0$ $\land$ Not($sub_p$)}) = $sat$}
    \State {input $FSM$ is unsafe ... abort}
\EndIf
\While{$k < 2^{n}$}
    \State{ $run_k$ = $T$}
    \For {\texttt{$i$ in $n$}}
        \State{$run_k$ = substitute($run_k$, ($X$[i],$V$[ n * {(k-1)+i]}),(Y[i],$V$[n* {k+i} ]))}
    \EndFor
    \State{$run_k$ = \text{Run[$k$ - 1] $\land$ $run_k$}}
    \State{$subp_k$ = $P$}
    \For {\texttt{$i$ in $n$}}
        \State{$subp_k$ = substitute($subp_k$, ($X$[i],$V$[n*{k+i}]))}
    \EndFor
    \If{(\text{$run_k$ $\land$ Not($subp_k$)}) == $sat$}
        \State{$FSM$ is unsafe ... abort}
    
    \Else
        \State{all states at distance $i$ are $P$-states}
    \State{$k$ = $k$ + 1}
    \EndIf
\EndWhile

\end{algorithmic}

\end{algorithm}
\section{Interpolation Technique}
This technique was first introduced by Kenneth L McMillan~\cite{mcmillan2003interpolation}. In this section we first discuss the flaws in BMC and how interpolation can be used to iron out these problems. Thereafter interpolation-based model checking algorithm is described in detail and later we compare these techniques on various examples.

In BMC, the output is only in the Boolean form, whether it is satisfiable or unsatisfiable but with interpolation, we get a proof of unsatisfiability and from that, we can create an interpolant which is used to prove that all reachable states are safe. As we have seen in the BMC algorithm we can only increment $k$ by 1, but in interpolation, we are able to increment $k$ by any number, as mentioned in next sections. Interpolation is an efficient methodology to enhance BMC towards the model checking of infinite-state systems and verification of safety and liveness properties. We give a brief introduction to Craig's interpolation theorem in the next section.

\subsection{Craig's Interpolant}
Craig's interpolation theorem states that for a contradiction \text{$A$ $\land$ $B$}, a Craig interpolant is a formula $P$\cite{mcmillan2005applications}, such that
\begin{itemize}[topsep=1pt,itemsep=1pt,partopsep=1pt, parsep=1pt]
    \item $A$ implies $P$
    \item \text{$P$ $\land$ $B$} is a contradiction
    \item All non-logical symbols of $P$ occur in $A$ as well as $B$.
\end{itemize}
%
%We can only create an interpolant if we have a proof by resolution that two clauses $A$ and $B$ are unsatisfiable and after creating an interpolant $P$, we can prove that $P$ is a Craig interpolant by satisfying the above three conditions.
%
%Suppose we have an initial state, transition relation, and property using that we can create two clauses $A$ and $B$. After receiving $A$ and $B$, we have passed these variables in the SAT solver and if it returns unsatisfiable then only we will proceed further. Suppose it returns unsat, then we will move forward towards the creation of interpolant $P$.  
An interpolant can be computed from the proof of unsatisfiability of \text{$A$ $\land$ $B$}. For more information about efficient interpolation computation, visit\cite{pudlak1997lower}. In this report, we have chosen a basic algorithm for interpolant computation.
\begin{algorithm}[H]
\caption{Computing Interpolant}
\label{alg:cap}
\begin{algorithmic}
\Ensure $Var = Var(A) - Var(B)$
\For{\texttt{$x$ in $Var$}}
    \State \texttt{$P$ = $A$[True|$x$] || $A$[False|$x$]}
\EndFor
\State {$P$ = simplify($P$)}
\end{algorithmic}
\end{algorithm}   

We consider the tuple ($VarA$, $VarB$) which store the list of states contained in clauses $A$ and $B$ respectively. Then, we take the difference between the two lists and store them in $Var$. After that, for every item $x$ in list $Var$, first, we replace True with $x$ in the $A$ clause and then replace False with $x$ and then do the OR operation. We repeat the same procedure in every iteration. After that to remove the redundant variables, we simplify our interpolant using simplify() which is an inbuilt function in the $Z3$ theorem prover\cite{moura2008z3}.

\subsection{Algorithm}
Given the original BMC formula for some $k$, we divide it into two clauses $A$ and $B$. '$A$' clause stores the initial state and first transition state whereas $B$ consists of the remaining transition state and property condition. Then an interpolant is constructed which is an over approximated image operator as the interpolant depends only on the $A$ clause. 

Let us consider the complete BMC formula \cite{mcmillan2003interpolation}:
    \[I(S_0) \land \bigwedge\limits_{i=0}^{k-1} T(S_i, S_{i+1})\land \bigvee\limits_{i=j}^{k} \neg P(S_k)\]
The $A$ and $B$ formulas in this case are:
\begin{flalign}
    \text{A\hspace{5pt} = \hspace{5pt} I($S_0$) \hspace{5pt} $\land$ T($S_0$, $S_1$)}
\end{flalign}
\begin{flalign}
    B\hspace{5pt} = \bigwedge\limits_{i=0}^{k-1} T(S_i, S_{i+1})\land \bigvee\limits_{i=j}^{k} \neg P(S_k)
\end{flalign}    
We have to introduce new indexed variables for each run which means we have to replace our state variables with path variables. Here \text{$S_i$ = {$s_{i1}$ ,$s_{i2}$, $\cdots$, $s_{in}$}}. 

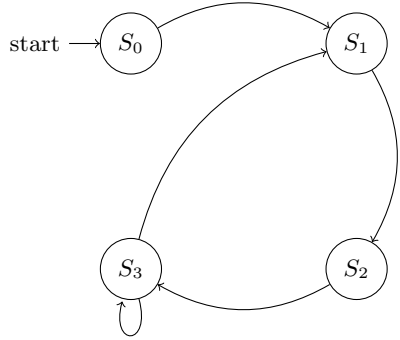
\begin{figure}%[!h]
\centering
\input{fig3}    
\caption{Finite State Machine with $4$ states}
\label{fig:fsm1interpolant}
\end{figure}

We now consider a finite state machine Figure.~\ref{fig:fsm1interpolant} to better understand interpolant computation.
%
%
%\vspace{50pt}

%\vspace{20pt}

For representing four states we need two lists which contains two Boolean variables (let's say $X$ and $Y$).
\begin{description}
\item $S_0$ is represented by \text{$\neg$$X_1$ $\land$ $\neg$$X_0$}
\item $S_1$ is represented by \text{$\neg$$X_1$ $\land$ $X_0$}
\item $S_2$ is represented by \text{$X_1$ $\land$ $\neg$$X_0$}
\item $S_3$ is represented by \text{$X_1$ $\land$ $X_0$}
\end{description}
Suppose the property to be checked is $P = X_1 \land X_0$. Initially, we consider $k = 1$. First we construct $A$ and $B$ formulas:
\begin{itemize}
    \item \text{A = $S0_0$ $\land$ ($S0_0$ $\implies$ $S1_1$) $\land$ ($S1_0$ $\implies$ $S2_1$) $\land$ ($S2_0$ $\implies$ $S3_1$) $\land$ ($S3_0$ $\implies$ ( $S3_1$ $\lor$ $S1_1$) ) }
    \item \text{B = $\neg$ $P_1$}
\end{itemize}
where $S0_0 = \neg X_{01} \land \neg X_{00}$, $S1_1 = \neg X_{11} \land X_{10}$ etc.
Using the above algorithm, we compute interpolant $ITP  = \neg X_{11} \land X_{10}$ which, essentially, represents state $S1_1$.

As shown above $ITP$ is the over-approximation of the forward image of the initial state (I). Here, initially, we take $k$ = 1, as we increase $k$, in each iteration, we will be able to compute the over-approximation of all reachable states. In BMC, we have to calculate the exact image operator but in the interpolation technique, we are using the over-approximation technique. With an increase in $k$, we will, at last, find a true counterexample or we can conclude that system is $P$-safe. If we increment $k$  by a too-small number, it will take more time to reach the conclusion or if we increment $k$ by a large number, it is possible that it will give a wrong result.

In algorithm 3, the create\_A\_and\_B($Q$, $T$, Not($P$)) function takes transition relation and property and returns $A$ and $B$ clauses which have been discussed earlier. createInterpolant($A$, $B$) takes $A$ and $B$ clauses and returns the interpolant which has been created from Algorithm 2.

\begin{algorithm}%[H]
\caption{Property Checking Procedure}\label{alg:check}
\begin{algorithmic}
\Procedure{PropertyChecking}{$prevK$, $K$}
\If{(\text{$I$ $\land$ Not(P)}) = sat}
    \State{\texttt{Property not satisfied}}
    \State{\texttt{abort}}
\EndIf
\State{$Q$ = $I$}
\While{$True$}
    \State{$A$,$B$ = $create\_A\_and\_B(Q,T, Not(P))$}
    \If{\text{$A$ $\land$ $B$} = sat}
    \If{$Q$ == $I$}
        \State{\texttt{Property not satisfied}}
        \State{\texttt{abort}}
    \Else
    \State{PropertyChecking($k$, $k + n$)}
    \EndIf
    \Else
        \State{$ITP$ = createInterpolant($A$,$B$)}
        \If{Not(Implies($ITP$, $Q$))}
            \State{Property holds}
            \State{abort}
        \EndIf
        \State{$Q$ = $Q$ $\lor$ $ITP$}
    \EndIf
\EndWhile
\EndProcedure
\end{algorithmic}
\end{algorithm}

\section{Comparison between $k$-induction and interpolation}
In this section, we consider two finite state machines and perform model checking on them using the techniques described above: $k
$-induction and interpolation, to illustrate the advantages and disadvantages of these algorithms.

\subsection*{Example $1$}\label{sec:eg1}
\begin{figure}[!ht]
\centering
\input{fig1}    
\caption{Finite State Machine with $4$ states}
\label{fig:machine1}
\end{figure}
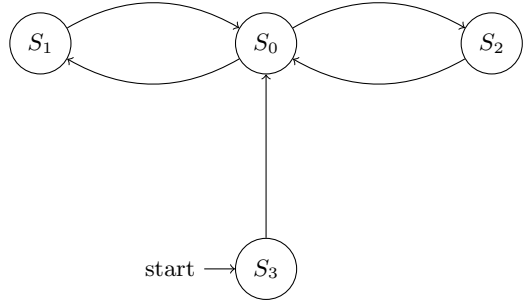

Now, we will show how BMC with $k$-induction works with an example from Figure.~\ref{fig:machine1}.

%\vspace{50pt}

%\vspace{20pt}
The initial state is $S_3$, and the property that we are checking is\text{ P = X[1] $\oplus$ X[0]} which means it should be satisfied in states $S_1$ and $S_2$ and the $FSM$ is $P$-safe for states $S_0$ and $S_3$. For each $k$ we have a $run_k$ and $subp_k$. These variables will then be passed into a solver and the result is shown after solving $SAT$ queries. 

\begin{table}[!ht]
    \centering
    \begin{tabular}{|c|l|c|c|}
        \hline
        k & $run_k$ & $subp_{k}$ & result \\ [0.5ex]
        \hline & & & \\ [-0.5ex]
        k = 0 & $run_0: S3_0$ & P = $X0_0$ $\oplus$ $X0_1$ & UNSAT\\
        \hline & & & \\[-0.5ex]
    
        k = 1 & $run_1: run_0$ $\land$ $S3_0$
        
        $\implies$ $S0_1$ $\land$ $S0_0$
        
        $\implies$ & & \\ & &  P = $X1_1$ $\oplus$ $X1_0$ & UNSAT\\
        
        &    ( $S1_1$ $\lor$ $S2_1$)$\land$ $S1_0$ $\implies$  $S0_1$ $\land$ $S2_0$ $\implies$  $S0_1$  & & \\
        
        \hline & & & \\[-0.5ex]
        k = 2 & $run_2: run_0$ $\land$ $run_1$ $\land$ $S3_1$ 
        
        $\implies$ $S0_2$ $\land$ $S0_1$ $\implies$ & &
        \\ & & P = $X2_1$ $\oplus$ $X2_0$ & SAT \\
        &
        ( $S1_2$ $\lor$ $S2_2$)$\land$ $S1_1$ $\implies$  $S0_2$ $\land$ $S2_1$ $\implies$  $S0_2$  & & \\ 
        \hline
    \end{tabular}
    \caption{BMC with $k$-induction }
    \label{tab:table1iteration}
\end{table}    

With $k$ = 1, all states reachable in one transition is $S_0$. and the result is unsatisfiable. Hence, we increment to $k$ = 2, where all states reachable in two transitions are {$S_0$, $S_1$, $S_2$} and property is satisfiable for $S_1$ and $S_2$ so it returns satisfiable and we can conclude that FSM is not $P$-safe and then we terminate the program.

We discuss the same example using interpolation-based model checking.
Here we have to keep track of previous and current $k$ variables. we can increase k by any amount but for this example, we increment it by $n$, where $n$ is the number of Boolean variables required to represent all states in FSM. So, initially, we pass our previous and current $k$ as $1$ and $1$ respectively.

\begin{table}[!ht]
    \centering
    \begin{tabular}{|c|c|c|c|c|}
        \hline
        k & Queries & Interpolant & Q & Result\\[0.5ex]
        \hline & & & & \\[-1.5ex]
        
        k = 1 & \text{$A$ $\land$ $B$}  & $S0_1$ & & UNSAT\\
        & \text{$\neg$($P$ $\implies$ $Q$)} & & \text{$S3$ $\lor$ $S1$}& SAT \\
        k = 1 & \text{$A$ $\land$ $B$} & & & SAT\\
        k = 3 & \text{$A$ $\land$ $B$} & & & SAT\\[1ex]
        \hline
    \end{tabular}
    \caption{BMC with Interpolation}
    \label{tab:interpolation}
\end{table}

Here, all the iterations are explained in detail,\\
For $k$ = 1,\\
First, we check whether \text{$I$ $\land$ $\neg$($P$)} is unsatisfiable, if it is satisfiable then FSM does not satisfy the property and we terminate the program.
In the above example, if the output is unsatisfiable, now we create $A$ and $B$ variables and check \text{$A$ $\land$ $B$}. The output is unsatisfiable. So it satisfies the property at $k$ = 1. We then created the interpolant, $P$, as you can see from the table $2$, which represents ($S0_1$) which is the next state from the initial state. Then, after replacing path variables with state variables, and storing them in R1, the solver checks \text{$\neg$($R1$ $\implies$ $Q$)}.
There are two options if it is unsatisfiable then property holds for all reachable states, else we will update our $R$ and do the next iteration.\\
So,  it again creates $A$ and $B$ clauses and checks \text{$A$ $\land$ $B$}, now the output is satisfiable. Now we have another condition, if \text{$Q$ == $I$}, we terminate the program and concludes that the system does not satisfy the property. But this is not the case here, so $k$ is incremented by $n$ which in this case is $2$ because the number of states is $4$.

For k = $3$,\\
After checking \text{$A$ $\land$ $B$}, which returns satisfiable, we check if \text{$Q$ == $I$}, which is true here. So we terminate the program and conclude that the $FSM$ does not satisfy the property for\text{ 1 $<$ $k$ $\le$ 3}.

\subsection*{Example $2$}\label{sec:eg2}
\begin{figure}%[!ht]
\centering
\input{fig2}    
\caption{Finite State Machine with $6$ states}
\label{fig:machine2}
\end{figure}
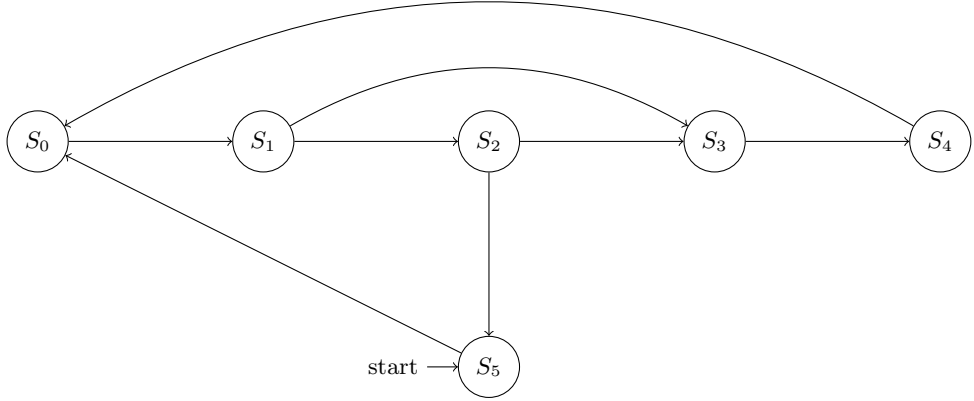

Now we look into another example as given in Figure.~\ref{fig:machine2} where property holds for all reachable states and check how these two algorithms perform.
%\vspace{50pt}

%\vspace{20pt}
Here the initial state is $S_5$ and the property that we are checking is 
Prop = \text{$X2$ $\land$ $X1$ $\land$ $X0$} which does not satisfy any of the reachable states. So the system will be property safe. Let's run our example with $BMC$ with the $k$-induction program.

\begin{table}%[h!]
    \centering
    \begin{tabular}{|c|c|c|c|}
        \hline
        k & $run_k$ & $subp_k$ & result \\ [0.5ex]
        \hline
        k = 0 & $S5_0$ & \text{ P = $X0_0$ $\land$ $X0_1$ $\land$ $X0_2$} & UNSAT\\[0.5ex]
        \hline
    
        k = 1 & \text{$run_0$ $\land$ $T_0$} & \text{P = $X1_0$ $\land$ $X1_1$ $\land$ $X1_2$} & UNSAT \\ [0.5ex]
        \hline 
        k = 2 & \text{$run_1$ $\land$ $T_1$} & \text{P = $X2_0$ $\land$ $X2_1$ $\land$ $X2_2$} & UNSAT \\ [0.5ex]
        \hline
        k = 3 & \text{$run_2$ $\land$ $T_2$} & \text{P = $X3_0$ $\land$ $X3_1$ $\land$ $X3_2$} & UNSAT \\ [0.5ex]
        \hline
        k = 4 & \text{$run_3$ $\land$ $T_3$} & \text{P = $X4_0$ $\land$ $X4_1$ $\land$ $X5_2$} & UNSAT \\ [0.5ex]
        \hline
        k = 5 & \text{$run_4$ $\land$ $T_4$} & \text{P = $X5_0$ $\land$ $X5_1$ $\land$ $X5_2$} & UNSAT \\ [0.5ex]
        \hline
        k = 6 & \text{$run_5$ $\land$ $T_5$} & \text{P = $X6_0$ $\land$ $X6_1$ $\land$ $X6_2$} & UNSAT \\ [0.5ex]
        \hline
        k = 7 & \text{$run_6$ $\land$ $T_6$} & \text{P = $X7_0$ $\land$ $X7_1$ $\land$ $X7_2$} & UNSAT \\ [0.5ex]
        \hline
    \end{tabular}
    \caption{BMC with k-induction }
    \label{tab:table1}
\end{table}  
After every result which is unsatisfiable, it reiterates till \text{$k < 2^{n}$}, and does the same checks. After $k$ = 7 our program will end and it will conclude that the system is P-safe. Here, $T_i$ = ( $T_0$, $T_1$, $T_2$, $\cdots$ ) represents the transition relation for each run, $k$.

\begin{table}%[h!]
    \centering
    \begin{tabular}{|c|c|c|c|c|c|}
        \hline
        k & i & Queries & Interpolant($P_i$) & $Q_i$ & Result\\[0.5ex]
         \hline
        k = 1 & 0 & \text{$A$ $\land$ $B$}  & $S0_1$ & & UNSAT\\
        & & \text{$\neg$($P$ $\implies$ $Q$)} & & \text{$S0$ $\lor$ $S5$}& SAT \\
        k = 1 & 1 & \text{$A$ $\land$ $B$}  & \text{$P_0$ $\lor$ $S1_1$} & & UNSAT\\
        & & \text{$\neg$($P$ $\implies$ $Q$)} & & \text{$Q_0$ $\lor$ $S1$}& SAT \\
        k = 1 & 2 & \text{$A$ $\land$ $B$}  & \text{$P_1$ $\lor$ $S2_1$ $\lor$ $S3_1$} & & UNSAT\\
        & & \text{$\neg$($P$ $\implies$ $Q$)} & & \text{$Q_1$ $\lor$ $S2$ $\lor$ $S3$}& SAT \\
        k = 1 & 3 & \text{$A$ $\land$ $B$}  & \text{$P_2$ $\lor$ $S4_1$ $\lor$ $S5_1$} & & UNSAT\\
        & & \text{$\neg$($P$ $\implies$ $Q$)} & & \text{$Q_2$ $\lor$ $S4$ }& SAT \\
        k = 1 & 4 & \text{$A$ $\land$ $B$}  & \text{$P_3$ } & & UNSAT\\
        & &  \text{$\neg$($P$ $\implies$ $Q$)}& & & UNSAT \\[1ex]
        \hline        
        
    \end{tabular}
    \caption{Interpolation}
    \label{tab:my_label}
\end{table}
At the end if result for \text{$\neg$($P$ $\implies$ $B$)} becomes unsat we can conclude that property holds for all reachable states.
\section{Implementation}
We need a SAT/SMT solver in the model checking algorithms presented in this report. We have used $Z3$ theorem prover (constraint solver) which is an efficient and open source SAT/SMT solver. We use ANTLR (Another Tool for Language Recognition) for input preprocessing as with ANTLR we can define our own input format and process it easily. In the next sections, we have introduced these tools and their usage in our implementation.

\subsection{Z3 Theorem Prover}
Z3 is used in hardware and software verification, constraint solving, and testing. Let us see a few examples to better understand its working. Here, we show how it is used for constraint solving and simplification of a Boolean expression.

\begin{Verbatim}[commandchars=\\\{\}]
    a = Int('a')
    b = Int('b')
    solve(a > 7, b >= 10, a + 2*b == 28)
    \color{red}{\textbf{OUTPUT}}
    [b = 10, a = 8]
\end{Verbatim}

In case a solution does not exist it returns ``No Solution''. We can also simplify the complex Boolean expression. 
\begin{Verbatim}[commandchars=\\\{\}]
    a = Int('a')
    b = Int('b')
    simplify(a**2 + 2*a**2 + b**2 - b**2 + 5)
    \color{red}{\textbf{OUTPUT}}
    5 + 3·a2
\end{Verbatim}
We have used $simplify()$ function in our implementation for simplification of interpolants generated in the course of execution of the algorithm because the interpolant produces redundant expressions. 

Now we look at $prove()$ function which prove a given statement by checking whether the statement negation is unsatisfiable or not.
\begin{Verbatim}[commandchars=\\\{\}]
    a = Bool("a")
    b = Bool("b")
    statement = Xor(a,b)
    prove(statement)
    \color{red}{\textbf{OUTPUT}}
    counterexample
    [b = False, a = False]
\end{Verbatim}
It gives a counterexample along with the values when it is not able to prove the given statement. Next, an example using $Solver()$ is demonstrated.
\begin{Verbatim}[commandchars=\\\{\}]
    a = Bool("a")
    b = Bool("b")
    s = Solver()
    s.add(Implies(a,b))
    s.check()
\end{Verbatim}
It returns $sat$ because for $a$ = True and $b$ = True a solution exists. $Solver()$ creates a general purpose solver and we can add constraints in the $Solver()$ using $add()$. For solving the constraints we use $check()$ which can return any of the three possible outputs. $sat$ if there exists a solution, $unsat$ if no solution exists and $unknown$ if $Solver()$ is not able to solve given constraints.

\subsection{ANTLR}
We have used ANTLR (Another Tool for Language Recognition) for input preprocessing of finite state machines. ANTLR is a parser generator. For getting a parse tree, first, we have to write a lexer and parser grammar then after running ANTLR it generates a lexer and parser in a targetted programming language. We then pass our finite state machine input in propositional formulas to obtain a parse tree. The interested reader may refer to the book~\cite{parr2013definitive} for a detailed study of ANTLR.
%\vspace{50pt}
\begin{figure}[!ht]
\centering
\input{fig3}    
\caption{Finite State Machine with $4$ states}
\label{fig:machine3}
\end{figure}
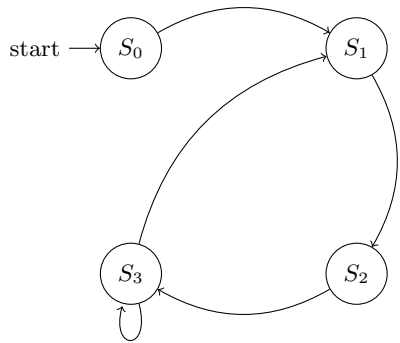

%\vspace{20pt}
For a finite state machine with 4 states given in Figure.~\ref{fig:machine3}, the following formulas have to be provided as input for model checking.  As one can see, it is not possible to manually write such formulas for more complex systems.

\begin{verbatim}
    I = And(Not(X[1]), Not(X[0]))
    T = And(Implies(And(Not(X[1]), Not(X[0])), And(Not(Y[1]), Y[0])),
    Implies(And(Not(X[1]), X[0]), And(Y[1], Not(Y[0]))),
    Implies(And(X[1], Not(X[0])), And(Y[1], Y[0])),
    Implies(And(X[1], X[0]), Or(And(Y[1], Y[0]),And(Not(Y[1]), Y[0]))))
    # define some properties.
    Prop = And(X[1], X[0])
\end{verbatim}
But with the help of ANTLR, we define our own input format, which is recognized using ANTLR, and feed it to our program. In this case, we can pass the input in this format:
%\textcolor{red}{ANTLR does not reduce the complexity of the input}
\begin{verbatim}
        SIZE : 2
        s0 : 00
        s1 : 01
        s2 : 10
        s3 : 11

        INIT(s0)
        NEXT(s0) := s1
        NEXT(s1) := s2
        NEXT(s2) := s3
        NEXT(s3) := s3
        NEXT(s3) := s1    
\end{verbatim}
The above input represents the FSM in Figure~\ref{fig:machine3}. Here SIZE refers to the number of Boolean variables required to create all states. INIT means the initial state of FSM and NEXT represents the transition state. NEXT($s0$) := $s1$, means there is a transition from $s0$ to $s1$. 

 After the parse tree is generated, we can walk the tree and there are two different ways, listeners and visitors. ANTLR generates a parse tree listener interface if we do not provide any specification. For trigger calls to listeners, ANTLR has $ParseTreeWalker$. It also generates a subclass that is unique for each grammar, $ExprListener$ for $Expr.g4$ grammar file, for example. For each rule, we have to provide enter and exit functions. So, the moment walker finds a node it encounters the node enter method and we can store or transform the data present at that node. The advantage of listener mechanisms is that you do not have to worry about the flow you just have to override the needed functions. 
 If you want more control over the nodes of your parse tree you can use an alternative mechanism called $visitor$. In this implementation, we have used the listener mechanism. We stored the data on each node and transform it into required formulas with the help of given information.

\section{Conclusion}
In this report we have discussed the interpolation and $k$-induction algorithms for model checking and their implementation in detail. First, a brief introduction on foundational topics viz. model checking, logic, safety, and liveness properties was given. Then, we discussed the bounded model checking algorithm, its pros and cons, and its implementation in detail. Later, an interpolation-based model checking algorithm was discussed. Next, we considered two examples to elucidate the internal working of these algorithms. Finally, we discussed the implementation, in particular the use of $Z3$ theorem prover and ANTLR in the project. 

We can not just conclude that the interpolation algorithm is better than BMC with $k$-induction. In example~\ref{sec:eg2}, where we selected a safe model, in the worst case, the upper bound can be the same as that of BMC with $k$-induction. But generally, the number of iterations is very small. In example~\ref{sec:eg1}, as we have shown, interpolant computation on an unsafe model is slower when compared to BMC with $k$-induction on the same model. Arguably, the importance of computing interpolants should not be ignored as there are other algorithms which have been developed using interpolants, like interpolation-sequence-based model checking\cite{vizel2009interpolation}, which is not in the scope of the report. 

%This report can help students to understand these model checking algorithms and this report also presents the approach and implementation of these algorithms. By going through the report, one can get an idea to choose which algorithm in which cases. Overall, after completing this report, you can get basic insights and it sets the foundation for you to dive deep into the model checking field. We have attached the source code and it's open for improvements.

\printbibliography

\end{document}

%% file: fig3.tex
\begin{tikzpicture}[->, node distance=3cm, auto]
\node[initial, state](A){$S_0$};
\node[state](B)[right of =A]{$S_1$};
\node[state](C)[below of =A]{$S_3$};
\node[state](D)[below of =B]{$S_2$};

\path (A) edge [bend left] (B)
    (B) edge [bend left] (D)
    (D) edge [bend left] (C)
    (C) edge [bend left] (B)
    (C) edge [loop below] (C);
\end{tikzpicture}

%% file: fig1.tex
\begin{tikzpicture}[->, node distance=3cm, auto]
\node[state](A){$S_1$};
\node[state](B)[right of =A]{$S_0$};
\node[state](C)[right of =B]{$S_2$};
\node[initial,state](D)[below of =B]{$S_3$};

\path (A) edge [bend left] (B)
    (B) edge [bend left] (C)
    (C) edge [bend left] (B)
    (B) edge [bend left] (A)
    (D) edge (B);
\end{tikzpicture}

%% file: fig2.tex
\begin{tikzpicture}[->, node distance=3cm, auto]
\node[state](A) {$S_0$};
\node[state](B)[right of =A] {$S_1$};
\node[state](C)[right of =B] {$S_2$};
\node[state](D)[right of =C] {$S_3$};
\node[state](E)[right of =D] {$S_4$};
\node[initial, initial text=start,state](F)[below of= C] {$S_5$};

\path (A) edge (B)
    (B) edge (C)
    (B) edge [bend left] (D)
    (C) edge (D)
    (C) edge (F)
    (D) edge (E)
    (E) edge [bend right] (A)
    (F) edge (A);

\end{tikzpicture}